\documentclass[aps,superscriptaddress,amsmath,amssymb,floatfix,twocolumn,showpacs,amsfonts,longbibliography,reprint]{revtex4-2}
\usepackage{times}
\usepackage[varg]{txfonts}
\usepackage{textcomp}
\usepackage{graphicx}
\usepackage{soul}
\usepackage[caption=false]{subfig}
\usepackage{color}
\usepackage[colorlinks=true,citecolor=blue,urlcolor=blue,linkcolor=blue,hyperindex]{hyperref}
\usepackage{braket}
\usepackage{float}
\usepackage{overpic}
\usepackage{bm}
\usepackage{array}
\usepackage{booktabs}
\usepackage{appendix}
\usepackage{tabularx}
\usepackage{cases}
\usepackage{multirow}
\usepackage{mathtools}
\usepackage{soul}
\usepackage{amsmath}
\usepackage{mathrsfs}
\usepackage{tikz}
\usepackage{tikz-3dplot}
\usetikzlibrary{positioning}
\usetikzlibrary{decorations.markings}
\usepackage[normalem]{ulem}
\usepackage{lineno}
\usepackage{graphicx}
\usepackage{dcolumn}
\usepackage{bm}

\newcommand{\datta}[1]{\textcolor{blue}{#1}}

\newcommand{\csbr}{\textrm{CrSBr}}

\begin{document}
\title{RIXS Identification of Optical Phonon-Spin Coupling Effects in CrSBr}

\author{Jayajeewana N. Ranhili}
\affiliation{Physics Department and Texas Center for Superconductivity, University of Houston, Houston, TX 77204, USA}

\author{Chamini S. Pathiraja}
\affiliation{Physics Department and Texas Center for Superconductivity, University of Houston, Houston, TX 77204, USA}

\author{Brody Brogdon}
\affiliation{Department of Physics and Biophysics, Augusta University, Augusta, GA 30912}

\author{John Cenker}
\author{Xiaodong Xu}

\affiliation{Department of Physics, University of Washington, Seattle, WA 98195}

\author{Daniel Chica}
\author{Xavier Roy}
\affiliation{Department of Chemistry, Columbia University, New York, NY 10027}

\author{Stefano Agrestini}
\affiliation{Diamond Light Source, Harwell Campus, Didcot, OX11 0DE, UK}

\author{Mirian Garcia-Fernandez}
\affiliation{Diamond Light Source, Harwell Campus, Didcot, OX11 0DE, UK}

\author{Ke-Jin Zhou}
\affiliation{Diamond Light Source, Harwell Campus, Didcot, OX11 0DE, UK}

\author{Yi-De Chuang}
\affiliation{Lawrence Berkeley National Laboratory, Berkeley, CA 94720, USA}

\author{Trinanjan Datta}
\email[Corresponding author:]{tdatta@augusta.edu}
\affiliation{Department of Physics and Biophysics, Augusta University, Augusta, GA 30912}

\author{Byron Freelon}
\email[Corresponding author:]{bkfreelon@uh.edu}
\affiliation{Physics Department and Texas Center for Superconductivity, University of Houston, Houston, TX 77204, USA}

\date{\today}

\begin{abstract}

Resonant inelastic x-ray scattering provides experimental signatures of spin-phonon coupling in CrSBr through temperature-dependent Cr $L$-edge spectra. Low-energy excitations are observed exclusively in the low-temperature antiferromagnetic phase as energy-loss features. A quasi-elastic peak at approximately 42 meV is observed under $\pi$-polarization. Density functional theory phonon-mode calculations identify these RIXS features as occurring within the same energy range as bond-bending optical phonon modes associated with distortions of the Cr--S--Cr network. The pronounced suppression of these low-energy excitations upon warming into the paramagnetic phase, together with their polarization dependence, the calculated phonon spectrum, and a spin-renormalized electron--phonon RIXS framework, indicates a strong interplay between magnetic correlations and lattice dynamics. While the loss features appear at energies characteristic of optical phonons, the significant overlap of the optical-phonon and magnon bands suggests that the temperature-dependent behavior should not be regarded as purely lattice-derived excitations. Instead, the room-temperature suppression of the low-energy RIXS peaks is explained in terms of a spin--phonon coupling effect on the $L$-edge electron--phonon RIXS mechanism. The interpretation is supported by the combined experimental observations, phonon calculations, and theoretical modeling, rather than by temperature contrast alone. These results support spin--phonon coupling as a plausible and consistent interpretation of the observed temperature-dependent RIXS response and demonstrate that magnetic order can strongly influence phonon-related spectral weight in the RIXS spectrum.
\end{abstract}

\maketitle

{\flushleft \textbf{Introduction}}

{\flushleft Quantum} materials in which lattice, orbital symmetry, electronic charge, and electronic spin are strongly coupled are especially promising~\cite{Tokura2017QuantumMaterials,Keimer2017PhysicsQuantumMaterials,Basov2017QuantumMaterials,Dagotto2005Complexity,Khomskii2014TransitionMetalCompounds,Basov2017QuantumMaterials}. Exploiting these intertwined degrees of freedom will lead to the creation of novel beyond-silicon electronic devices. Among such materials, the two-dimensional (2D) van der Waals (vdW) magnets have become a major research focus \cite{wang2022magnetic,gibertini2019magnetic,gong2019two,burch2018magnetism,pathiraja2025electronic}. The magnetic ordering in these materials can be easily tuned through temperature, electric and magnetic fields, pressure, and strain~\cite{pham20222d,forg2021moire,barik2021two}. Since the discovery of 2D magnetism~\cite{huang2017layer}, chromium-based vdW magnets have been specifically explored because they exhibit strong coupling that results in magneto-optical effects, magnetoresistance~\cite{cenker2021direct,wang2018very,peng2022multiwavelength,zhang2019direct,fkas2025direct,wu2024magneto}, and tunable photoluminescence, making them attractive for spintronic and optoelectronic applications~\cite{ningrum2020recent,cortie2020two,liu2020spintronics}. Despite these advances, direct spectroscopic identification of spin-phonon coupling effects and its impact on low-energy excitations is limited, particularly in momentum and energy resolved probes.

The 2D vdW magnet chromium sulfur bromide (\csbr) has emerged as a prominent material for studying coupled magnon–photon–phonon interactions that are tunable based on optical and other external stimuli~\cite{tang2025coherent,ranhili2025ultrafast,shen2025orthogonal}. This layered vdW magnet stands out due to its rich landscape of quasiparticle excitations spanning from the meV to the eV energy scale~\cite{ziebel2024crsbr,tschudin2024imaging,han2025exciton}. The low-dimensional magnetic ordering in \csbr~couples strongly to charge, lattice, and excitonic degrees of freedom~\cite{bae2022exciton,datta2025magnon,lin2024probing,ranhili2025ultrafast}, enabling the emergence of composite quasiparticles, such as magnon–polaron modes~\cite{simensen2019magnon,vaclavkova2021magnon,godejohann2020magnon}. The magnetic ordering of \csbr~can be thermally driven. Near 146 K, the material develops an in-plane ferromagnetic order~\cite{rudenko2023dielectric,ziebel2024crsbr}.  Below approximately 132 K, weak antiferromagnetic coupling between the layers develops concomitantly with a strong intra-layer ferromagnetic interaction to stabilize a long-range $A$-type antiferromagnetic order ~\cite{goser1990magnetic,rudenko2023dielectric,ziebel2024crsbr}. The \csbr~optical band gap is in the range of 1.5 eV to 1.8 eV~\cite{telford2020layered,ziebel2024crsbr, klein2023bulk}, thus making it a promising material for magnetic semiconductor applications~\cite{ziebel2024crsbr}. The orthorhombic bulk CrSBr lattice establishes connectivity via the Cr-S-Br octahedra, the quasi-1D magnetic exchange along the $a$-axis, and the local Cr$^{3+}$ ligand-field environment. See Supplementary Material Section~I for crystal structure details. These structural and electronic features determine the interplay of phonon, magnon, and electronic excitations which persist as the material is exfoliated down to the mono- or bi- layer version from its bulk \cite{li2025stacking,liu2024intralayer,antoniazzi2025magneto,qian2023anisotropic,lee2021magnetic}. 

Recent optical spectroscopy studies have established CrSBr as a prototypical platform for investigating coupled spin, lattice, and electronic degrees of freedom. Temperature-dependent Raman measurements demonstrated that magnetic ordering induces pronounced changes in phonon intensities, activates additional phonon modes, and generates higher-order scattering processes, providing direct evidence for strong spin-phonon coupling in bulk CrSBr.~\cite{pawbake2023raman} Additional Raman investigations revealed distinct spectroscopic signatures associated with the multiple magnetic phases of CrSBr and showed that lattice vibrations are highly sensitive to the evolution of magnetic order.~\cite{Wdowik_2025}  More recently, polarization-resolved Raman studies uncovered excitation-energy-dependent polarization switching of Raman-active modes and strong anisotropic electron-phonon coupling linked to the orthorhombic crystal structure.~\cite{mondal2025} Complementary optical measurements identified strong exciton-phonon coupling that evolves across the magnetic transition and serves as a sensitive indicator of magnetic ordering.~\cite{lin2024strong} Together with theoretical and experimental studies of magnetoelastic coupling~\cite{shen2025orthogonal}  and magnon-phonon interactions, these results establish that lattice excitations are strongly influenced by magnetic correlations in CrSBr.~\cite{shen2025orthogonal,ranhili2025ultrafast,datta2025magnon,Wdowik_2025} However, Raman spectroscopy does not directly probe the element-specific resonant intermediate states that govern x-ray scattering processes. Consequently, how spin-renormalized lattice dynamics manifest in Cr $L$-edge RIXS remains largely unexplored. Here, by combining temperature-dependent Cr $L$-edge RIXS measurements, density functional theory, and a spin-phonon renormalization framework, we extend the growing body of optical studies into the x-ray regime and demonstrate that magnetic correlations influence phonon spectral weight in the RIXS response.

From a spectroscopic perspective, bulk-sensitive soft x-ray techniques, especially Cr $L_{2,3}$-edge x-ray absorption spectroscopy (XAS) and resonant inelastic x-ray scattering (RIXS), provide detailed access to excitations that span low-energy (meV) to high-energy (eV) \cite{ament2011resonant,pathiraja2025electronic,pathiraja2025comparison}. In this investigation, Cr $L_{2,3}$-edge XAS and RIXS measurements were utilized to probe the low-energy excitation spectrum of \csbr. We uncover a distinct signature of spin–phonon coupling. Specifically, low-energy phonon excitations are clearly observed in the RIXS spectra at low temperature, but become strongly suppressed upon warming into the paramagnetic phase. Specifically, we observe the presence of low-energy phonon excitations in RIXS spectra at low temperature, which are strongly suppressed upon warming into the paramagnetic phase. This temperature dependent behavior is unexpected from a purely lattice-driven perspective, signaling the key role played by magnetic correlations. The observed behavior is explained within the framework of spin renormalized phonon frequencies which affects the electron-phonon coupled RIXS scattering profile. The experimental observation and physical explanation of this subtle temperature and spin-phonon coupled RIXS behavior at the Cr $L$-edge establishes a link between magnetic correlations and lattice excitations. The findings highlight the role RIXS can play as a sensitive probe of spin correlations on phonon dynamics. 

\begin{figure}[!ht]
     \centering
    \includegraphics[width=0.5\textwidth]{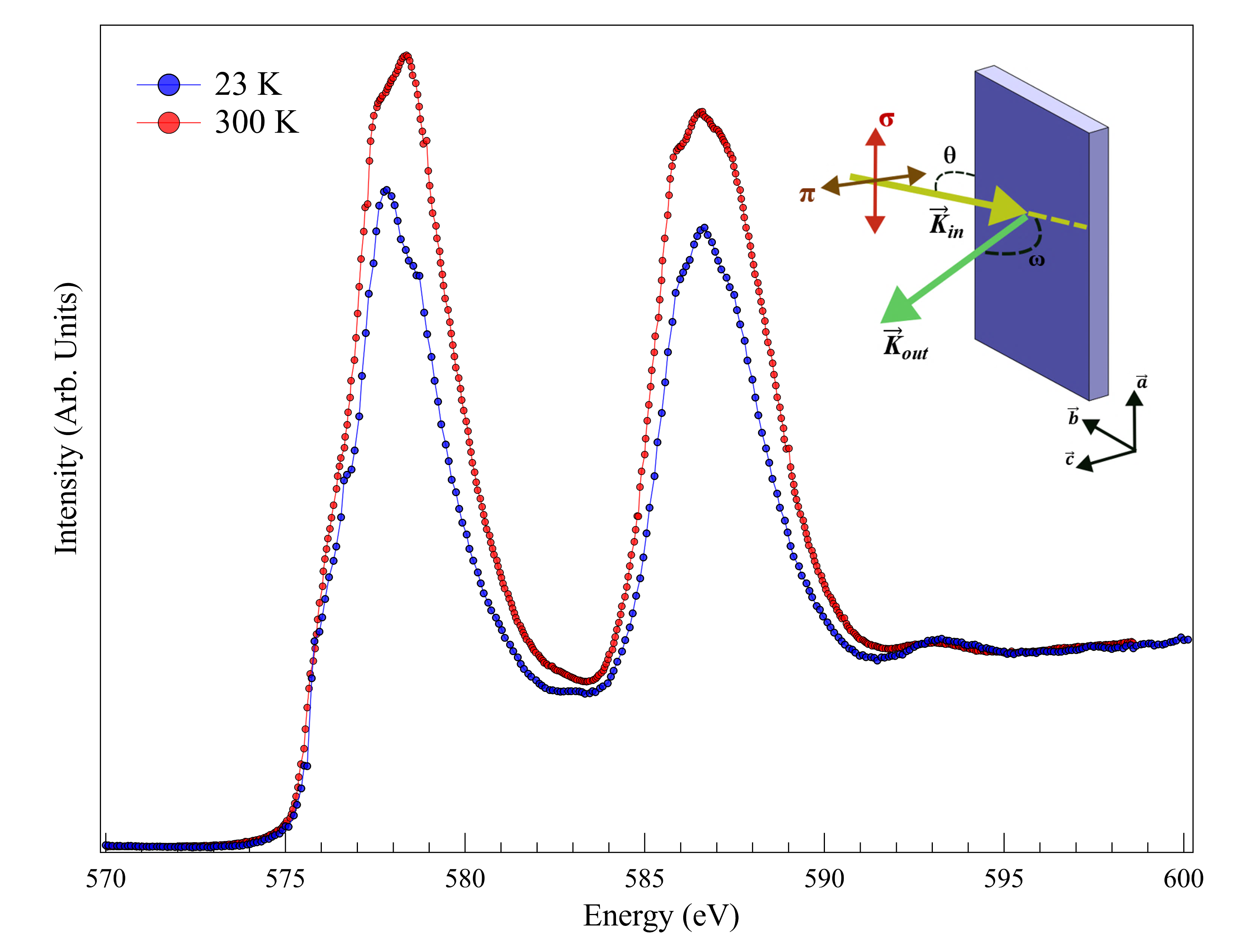}
    \caption{XAS spectra collected at 23 K (blue) and 300 K (red) under $\pi$-polarization in total electron
    yield (TEY) mode. The $L_3$ peak energy in both spectra is $\sim$577.8 eV. The inset shows the experimental geometry used at the I21 RIXS beamline to collect RIXS data. In $\pi$ ($\sigma$)-polarization geometry, the incident x-ray beam is polarized parallel (perpendicular) to the scattering plane. } 
    \label{XAS}
\end{figure}

{\flushleft \textbf{Results}}
{\flushleft \textbf{XAS measurements}}
{\flushleft Cr} $L$-edge XAS data were collected at x-ray incident angle $\theta$ of $80^\circ$ with respect to the surface of CrSBr. See the Figure~\ref{XAS} inset for RIXS experimental geometry details used at the I21 beamline. Data was taken in the energy range of 570 eV to 600 eV, spanning over the Cr $L_{2,3}$ edges, in total electron yield (TEY) mode. Figure~\ref{XAS} shows the XAS data at room-temperature (RT) and 23 K low-temperature (LT) phase using $\pi$-polarization. The $L_3$ peak energy at LT and RT is near 577.8 eV, and the pre-edge is close to 576.5 eV, which are closer to the previously reported results \cite{poree2025resonant,sears2025observation,hunault2018direct}. The RT XAS data was collected using a different grating with a slightly lower resolution of 28 meV. The XAS data collected under $\sigma$-polarization also shows similar peak energies (see Supplementary Figure S2). RIXS spectra were collected at the $L_3$ -edge peak energy, $L_3$ -1.5 eV, and $L_3$ +0.8 eV under $\pi$ and $\sigma$ polarizations at the same incident angle of $80^\circ$.

{\flushleft \textbf{RIXS measurements}}
\begin{figure*}[hbtp]
     \centering
    \includegraphics[width=1\textwidth]{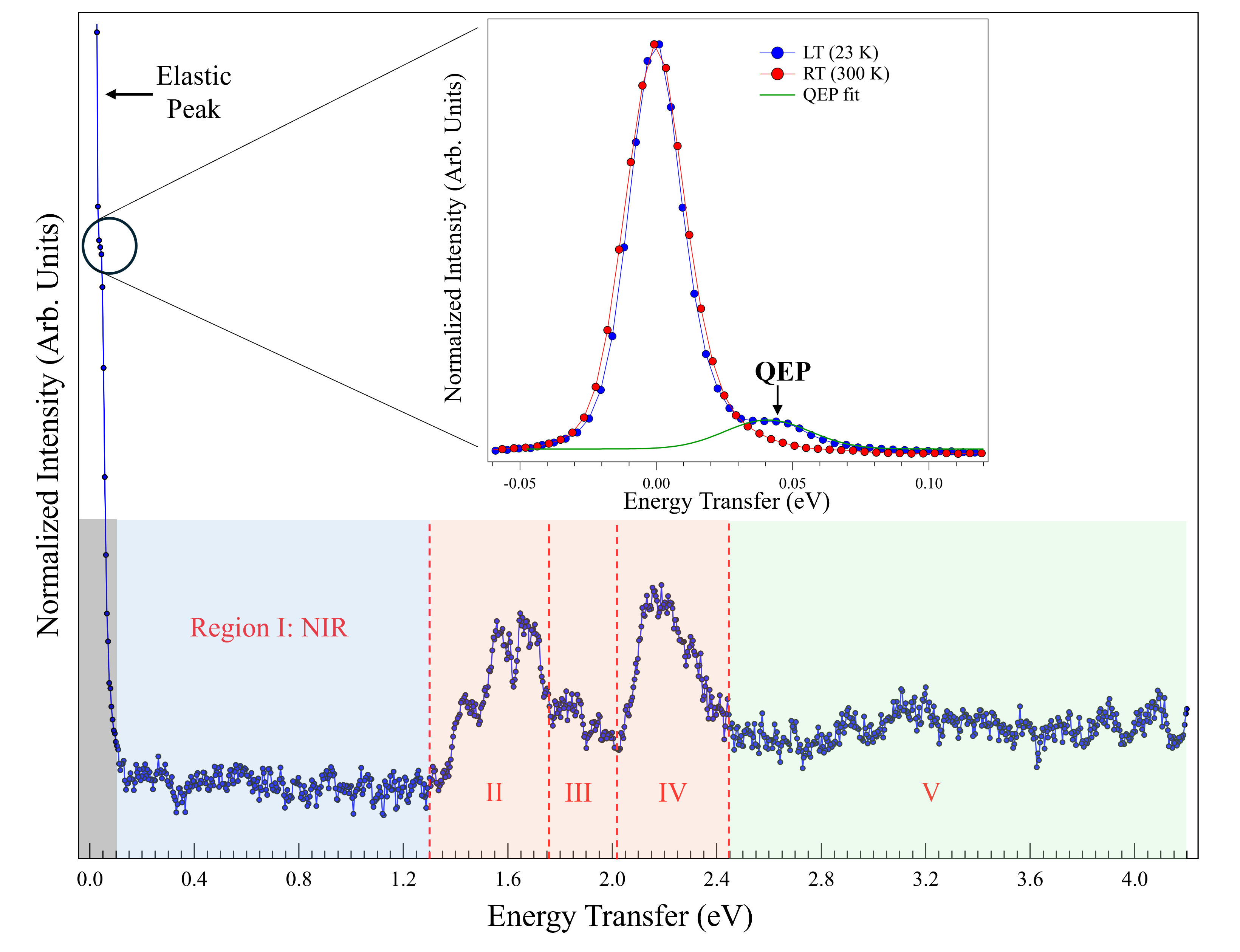}
    \caption{RIXS spectrum of bulk CrSBr collected under $\pi$-polarization at the $\Gamma$ point at 23 K (LT). The spectrum is normalized at the elastic peak (EP) and divided into several regions to highlight the various observed excitations. The inset shows the appearance of quasi-elastic peak (QEP) near to the EP in the low-temperature phase.}
    \label{label_Fig2}
\end{figure*}

{\flushleft Figure} \Ref{label_Fig2} shows elastic-peak normalized RIXS spectrum taken at 23 K under $\pi$-polarization at the $\Gamma$ point. The beam in- and out- angles are the same at the $\Gamma$ point, creating an in-plane momentum projection which is zero ($q = 0$). RIXS energy loss spectra can contain spectral weight relevant to a wide range of electronic excitations \cite{ament2011resonant, kotani2001resonant} as shown in Figure \Ref{label_Fig2}. For clarity, the spectral data has been demarcated by the elastic peak (EP), a quasi-elastic peak (QEP), and energy regions: I, II, III, IV, and V.

The EP, with spectral weight at 0 eV in Figure \ref{label_Fig2},  is a zero energy loss feature that occurs when the incident x-ray energy $E_{in}$ equals the scattered x-ray energy $E_{out}$. A QEP is observed as a small EP shoulder. A magnified view of the QEP region is shown as an inset of Figure~\ref{label_Fig2}. The inset shows both low and room temperature spectra collected at $q$ = 0 at the zone center. The energy of the QEP at LT under $\sigma$- polarization is $\sim$42.1 meV; the QEP feature is suppressed at 300 K. 
The RIXS QEP energy is close in value to optical phonon modes observed in several reports. Infrared spectroscopy studies found a spectral feature at 360 cm$^{-1}$ ($\approx$ 44 meV) that was attributed to a magnetic (magnon) excitation.~\cite{pawbake2023magneto,uykur2024phonon}  

The intrinsic linewidth of the QEP can be estimated by deconvolving the experimental resolution assuming a Gaussian instrumental response.~\cite{FumagalliPhysRevB.99.134517} The observed QEP width  $\Gamma_{\rm obs}$ $\sim$ 35 - 42 meV  with the experimental resolution giving a width $\Gamma_{\rm res}\approx24.5\ {\rm meV}$.  The intrinsic width of the peak can be calculated with the expression $\Gamma_{\rm intrinsic} \approx \sqrt{\Gamma_{\rm obs}^2-\Gamma_{\rm res}^2}$ which yields roughly 25 - 34 {\rm meV}. Such widths are too large for a simple optical phonons; therefore, the observed widths $\Gamma_{\rm obs}$ likely indicate the composite quasiparticle nature of the QEP and not the lifetime of a bare phonon. 

Density functional theory (DFT) computations, discussed below, indicate that phonon density of states is also clustered near the 40 meV (10 THz) region. Therefore the QEP in the RIXS data is energetically compatible with optical phonon modes that represent atomic lattice vibrational modes of CrSBr. However, the QEP has proximity to reported zone-center magnetic excitations ($\approx$ 44 meV). The two lines of evidence suggest that the QEP is in close energetic proximity both a zone-center optical magnon mode and a zone-center optical phonon mode. The presence of  phonon-magnon coupling~\cite{shen2025orthogonal,bae2022exciton} in CrSBr is well-established and suggests that the QEP may possess a mixed or coupled quasiparticle nature. Thus, coincidence of phonon and magnetic energy scales near ~42 – 44 meV suggests that the observed mode arises from a strongly coupled spin–lattice excitation rather than a purely phononic or magnetic origin. 

Region I includes near-infrared (NIR) energies and is labeled from $\sim$100 meV~-~1.3 eV. The spectral profile in this energy range shows negligible intensity. Region II to IV, which fall within the energies from 1.3 to 2.45 eV, contain $dd$ spectral peaks. However, region III contains spin-flipped $dd$ excitations that separate the low energy $dd$ from the high energy ones observed in the window II and IV \cite{sears2025observation}. The $dd$ excitations arise from local inter-orbital transitions within the crystal-field-split Cr 3$d$ manifold. These excitations are distinct from excitons, which correspond to bound electron-hole quasiparticles. In addition to the dd features, two peaks at 1.39 and 1.43 eV were observed in Region II and are assigned to the bright and dark excitons reported \cite{sears2025observation} previously for CrSBr using high-resolution Cr $L_3$-edge RIXS. The excitonic assignment is therefore based on the emerging CrSBr literature rather than on the present work alone. These two peaks appear prominently only in the low-temperature spectra. Therefore, the spectral weight at 1.39 eV and at 1.43 eV are assigned as bright \textbf{B} and dark \textbf{D} excitons, respectively. The observation of \textbf{B} and \textbf{D} high-energy features validates the experimental spectra (Figure \ref{label_Fig2}) of CrSBr~\cite{sears2025observation, wang2023magnetically}. The bright exciton utilizes a de-localized Wannier character arising from hybridization between the Cr $3d$ and S and Br $p$ states, these features were claimed to be the result of coupling to either phonon or polariton modes \cite{wilson2021interlayer,sears2025observation, wang2023magnetically}. Spectral features observed in region V from $\sim$2.45~-~4.0 eV are likely due ligand to metal charge transfer (LMCT) excitations. Such charge transfer would be due to S or Br ligand electron excitations into an empty or partially filled $d$ orbitals in Cr$^{3+}$ ion. A detailed assessment of this spectral weight warrants simulations but is beyond the scope of the present work.  

The focus in this paper is to identify the degrees of freedom which give rise to the low-temperature QEPs and to explain the physical origin of the RIXS spectrum temperature suppression. Below we provide a physical explanation of the QEP peaks using established theoretical models used to interpret RIXS signals from correlated electron systems.

\begin{figure*}[hbtp]
    \centering
    \includegraphics[scale=0.33]{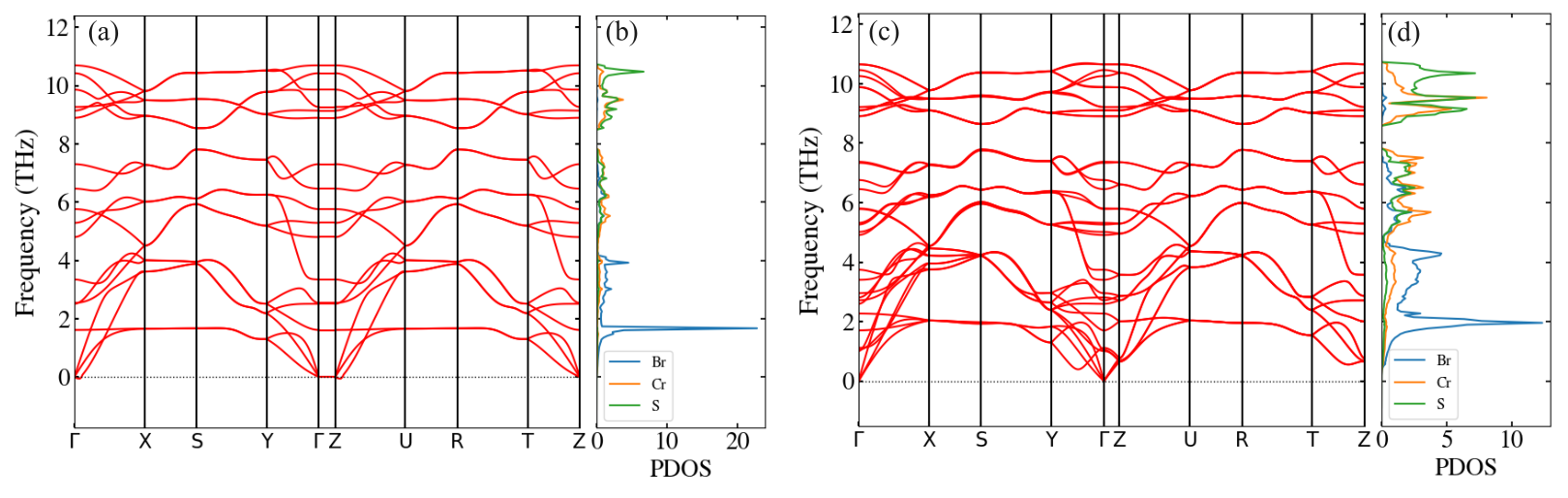}
    \caption{Density functional theory calculations. (a, b) Phonon dispersion and atom-projected phonon density of states (PDOS) for monolayer ferromagnetic configuration, (c,d) Phonon dispersion and atom-projected PDOS for bulk antiferromagnetic configuration. Note, 10 THz is equal to 41.36 meV.}
    \label{fig:fig7}
\end{figure*}
{\flushleft \textbf{Phonon Density Functional Theory}} {\flushleft The} RIXS data indicates the presence of distinctive low temperature peaks. The energy scale of these energy loss features hint towards the influence of phonon on the RIXS spectrum. However, their suppression at room temperature indicates that lattice vibrations alone cannot account for the observed suppression of intensity. This implies that additional mechanisms, beyond purely phononic degrees of freedom, contribute to the temperature-dependent RIXS response. In order to investigate this, we performed first-principles DFT phonon calculations to develop a spin-phonon coupling framework for CrSBr \cite{baltensperger1968influence,lockwood1988spin,wysocki2016magnetically}. The resulting insights were applied to analyzing the temperature-dependent electron-phonon-coupled RIXS response at the $L$-edge \cite{ament2011determining}. Bulk phonons including out-of-plane optical modes and interlayer shear or breathing modes are inherently three-dimensional and cannot be inferred solely from monolayer studies~\cite{pawbake2023raman,mondal2025raman,lin2024strong}. Furthermore, the magnon excitations are a manifestation of strong intralayer ferromagnetic exchange and weak interlayer antiferromagnetic coupling~\cite{lee2021magnetic,liu2024intralayer,ziebel2024crsbr}. These two facts are important in analyzing the interplay of spin-phonon coupling on the RIXS spectrum.


Density functional theory was used to compute CrSBr phonon dispersion bands, both for the monolayer and the bulk compound. We also calculated the atom-projected phonon density of states (PDOS). The results are displayed in Figure~\ref{fig:fig7}. The computations were performed using Quantum Espresso (QE) \cite{giannozzi2017advanced,giannozzi2009quantum} and the PHONOPY package \cite{phonopy-phono3py-JPCM,phonopy-phono3py-JPSJ}, see Methods for implementation details. The unit cell of the monolayer ferromagnetic CrSBr consists of six atoms, in which there are two Cr, two S, and two Br atoms. The monolayer is ferromagnetic while the bulk is antiferromagnetic. When constructing the bulk antiferromagnetic configuration for DFT, one must add an additional layer on top of the monolayer unit cell. Thus, we have a twelve atom unit cell for the bulk. The six atom bottom layer has spins oriented in a ferromagnetic pattern. The bulk antiferromagnetic structure was created with spins of opposite orientation between the mono-layers. As expected, based on the number of atoms in the unit cell, we note from Figs.~\ref{fig:fig7}(a) and \ref{fig:fig7}(c) that the monolayer (bulk) phonon band structure has 18 (36) phonon bands. There is a cluster of monolayer and bulk phonon bands in the vicinity of the energy where we notice the low temperature peaks in the RIXS spectrum of Figure~\ref{label_Fig2} (see figure inset). For the bulk compound, on which the experiment was done, DFT calculations imply that the optical phonon bands of interest range from 40.863 meV (9.881 THz) for the 31$^{\text{st}}$ band to 44.056 meV, 10.653 THz for the 36$^{\text{th}}$ band. These represent the six highest optic mode energies. For the monolayer configuration, the phonon bands range from 40.804 meV (10.430 THz) for the 16$^{\text{th}}$ band to 
44.250 meV (10.700) for the 18$^{\text{th}}$. 

The second and the fourth panel, Figures~\ref{fig:fig7}(b) and ~\ref{fig:fig7}(d), show the atom-projected phonon density of states (PDOS). The calculation shows an interesting pattern. In the acoustic region, 0 meV (0 THz) to 8.3 meV (2 THz), the PDOS indicates that the bromine (Br) atoms are the major contributors to generating lattice phonons. However, as we go higher up in energy towards the optic zone, this trend flips. At these optic energy levels, the chromium (Cr) and the sulfur (S) atoms are the major contributors. There is a tiny but negligible contribution from Br atoms in this regime. The monolayer and bulk PDOS suggest that Cr (and S) atoms in bulk have a significantly more phonon contribution than it's monolayer counterpart. We analyzed these optical phonon bands using the Phonon Website (\url{https://tinyurl.com/phononwebsite}) and SMODES~\cite{stokes2007isotropy}. This reveals details on the lattice motion of each species of atom in the unit cell and also classifies the irreducible representations. The lattice snapshots are displayed in Supplementary Figure~S4. The nature of the relative in- or out- of phase oscillations among similar species of atoms and the corresponding irreducible representation are summarized in Supplementary Table S1 (in Supplementary Section IV). Of the six top most optic bands, the lowest four (31 - 34) are bond-bending modes, see Supplementary Figures~S4(a)-(f). The top two (35 - 36) are bond-stretching modes, see Supplementary Figures~S4(g)-(i). The proximity of the low-temperature $\sim$43 meV peaks to the calculated optical phonon bands suggests that bond-bending 34$^{\text{th}}$ phonon optic mode involving the Cr–S–Cr network contribute significantly to the observed excitation.  
While the RIXS peak energies are in close proximity to the phonon band values, due to the interplay of multiple excitation modes (phonon and magnon) in \csbr, one should not associate the origin of these peaks exclusively with a pure phonon mode. Rather, as we discuss in the next paragraph (and in the supplementary), the low energy RIXS features are influenced by spin-phonon coupling which in turn \emph{influences} the electron-phonon coupling strength to yield the temperature dependent RIXS spectrum. While the location of the peak energies coincides with the optical phonon bands obtained from DFT computations, 
the suppression of the peaks with temperature variation requires further careful consideration. 

{\flushleft \textbf{Discussion}} {\flushleft A} central observation in the RIXS data of Figure~\ref{label_Fig2} is the suppression of the low energy RIXS peak with increasing temperature. This behavior is counter-intuitive, since a purely lattice-driven mechanism would predict an increase in phonon population, which in turn would have enhanced the RIXS spectral weight at RT. But upon cooling from room to low-temperature the compound has undergone several magnetic phase transitions from the paramagnetic phase, to a ferromagnetic phase, and then eventually transforms to an antiferromagnetic phase. Thus, in order to interpret the suppression of the RIXS peak, we need consider how phonon frequencies are affected by spin dynamics. To understand this behavior, we have formulated a spin-renormalized phonon frequency theory for CrSBr based upon a spin-phonon Heisenberg model Hamiltonian $H(Q)$, written as, \begin{equation}
\label{eq:hq}
H(Q) = \frac{P^2}{2M} + \frac{1}{2}\kappa_0 Q^2
\;-\;\sum_{\{ij\}} J_{ij}(Q)\, \mathbf S_i\!\cdot\!\mathbf S_j,
\end{equation} where $M$ is the mass of the atom, $P$ is the canonical momentum conjugate to $Q$ (single phonon coordinate), and $\kappa_0$ is the bare spring constant. The third term represents the Heisenberg type spin-phonon coupling Hamiltonian where $\mathbf S_i$ is the spin operator at site $i$. The phonon coordinate dependent exchange coupling is given by $J_{ij}$. The summation index pair $\{ij\}$, where $i$ and $j$ represent site indices, is over various nearest-neighbor $(nn)$ bonds. We consider the dominant intralayer exchange terms for \csbr~(up to the third neighbor). Further neighbor exchange or interlayer interactions are ignored since they are weaker in strength~\cite{Bo2023}. The theoretical formulation is detailed in Supplementary Section~V. 

To keep the discussion tractable, we briefly outline the main logic of the spin-phonon coupling formulation~\cite{spphn1,spphn2,BirolPhysRevB.93.134425}. First, we Taylor expand the exchange constant $J_{ij}(Q)$ about $Q=0$. Next, we collect the expansion terms that renormalize the phonon frequency to derive the renormalized phonon frequency $\omega^2_r$ expression (Equation S4 in Supplementary Material). Then, we apply the Goodenough-Kanamori-Anderson rules~\cite{GoodenoughPhysRev.100.564,Kanamori195987,AndersonPhysRev.115.2} to model the variation of the exchange parameter derivatives $K_l$, consider the mean field Heisenberg correlator $C_{ij}(T) \equiv\langle\mathbf{S_i}\cdot\mathbf{S_j}\rangle$, where $T$ is the temperature, to obtain the final expression for the renormalized phonon frequency as
\begin{equation}
\omega_{\mathrm{r}}^2 \approx \omega_0^2 - \mathrm{p}_lC_{l}, 
\end{equation}
in terms of the most dominant leading correction (see Supplementary Material for details). Here $\omega_0^2$ is the phonon frequency of the paramagnetic phase, $\mathrm{p}_l = K_l\lambda^2_l/M$ is the weight for the most dominant bond contribution (bond-bending), and $C_l$ is the intralayer correlator. Note, $\lambda_l$ is a geometric effect of the bond angle, see Supplementary Material Section V. Based on the Goodenough-Kanamori-Anderson rule (as reasoned in the Supplementary Material), $K_l > 0$ in \csbr~for the intralayer ferromagnetic channel. Furthermore, $C_l(T)>0$ for planar spin-spin correlation. Thus, $\Delta \omega^2(T) = \omega^{2}_r -\omega^2_0 = -(4/M) K_l\lambda^2_lC_l(T)$ is negative. The sign of $\lambda_l$ is irrelevant since it is squared. As temperature rises ferromagnetic correlations weaken, implying $C_l \rightarrow 0$. This leads to a hardening of the phonon frequency. Eventually, thermal fluctuations overpower the ferromagnetic exchange energy to drive the system to a paramagnetic phase, where $\omega_0$ is stabilized. But, the bare frequency is higher in value compared to the low temperature value where it was softened due to the negative contribution from $\Delta \omega^2(T)$. 

\begin{figure}[hbtp]
     \centering
    \includegraphics[width=0.5\textwidth]{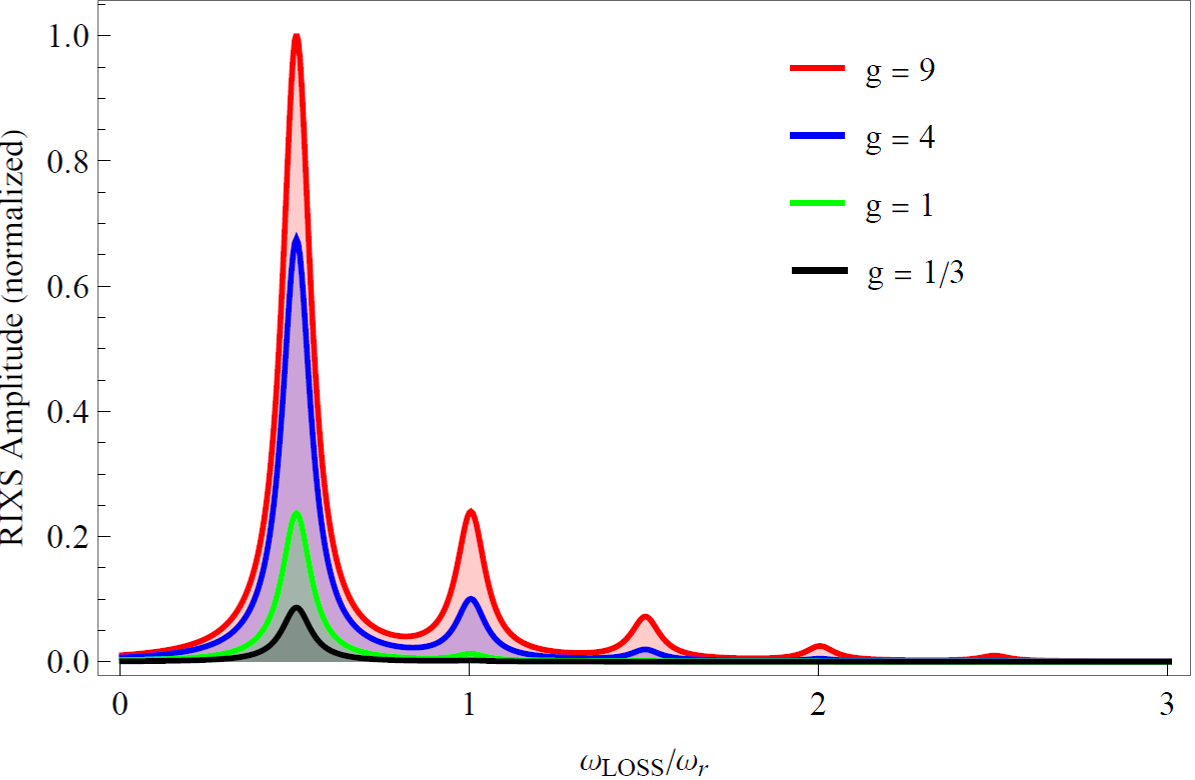}
    \caption{Normalized electron-phonon $L$- edge RIXS intensity. The electron-phonon coupling strength $g$ is inversely proportional to the square of the renormalized phonon frequency $1/\omega^2_r$. With increasing temperature the phonon frequency hardens thereby leading to a reduction in $g$, which in turn suppresses the RIXS spectral intensity. This trend is observed in the room temperature RIXS data of Figure~\ref{label_Fig2}.}
    \label{fig:fig3}
\end{figure}

Unlike conventional Raman spectroscopy, which primarily probes phonon eigenmodes through the optical polarizability response, phonon intensities in transition-metal L-edge RIXS are controlled by resonant electron-phonon scattering matrix elements associated with the intermediate electronic state. Consequently, the temperature dependence of a RIXS phonon feature cannot be inferred directly from Raman phonon shifts or linewidths, and comparatively small phonon renormalizations can produce much larger changes in the observed RIXS spectral weight. Based upon electron-phonon coupled RIXS theory~\cite{ament2011determining}, we can infer that the dimensionless electron-phonon coupling $g$, is inversely proportional to the square of the phonon frequency (that is $g\propto 1/\omega^2_r$). Also, utilizing the frequency hardening idea proposed in the previous paragraph, we can infer that the phonon frequency will vary based on temperature. Using these two facts as input, we have computed the electron-phonon coupled $L-$ edge RIXS spectrum~\cite{Ament2011}, influenced by spin-phonon coupling. The result shown in Figure~\ref{fig:fig3}, clearly indicates that when the phonon frequency is hardened (the lowest value of $g \propto 1/\omega^2_r$, the dimensionless electron-phonon coupling), the RIXS intensity contribution is suppressed. This is precisely the same behavior that is observed in the experimental data. This explains why the RIXS intensity is subdued with increasing temperature. Finally, note that although the low-temperature phase exhibits antiferromagnetic stacking, the associated interlayer exchange scale is sufficiently small that the corresponding spin correlations make a negligible contribution to the phonon self-energy, leaving the renormalization dominated by intralayer ferromagnetic correlations which produce an overall positive contribution. Thus, the interplay of spin-phonon coupling on RIXS is crucial in explaining the RIXS intensity suppression.

In summary, we demonstrate clear signatures of spin-phonon coupling in the temperature dependent CrSBr RIXS spectrum, while confirming the simultaneous presence of both bright and dark excitons in the same magnetic phase which has active spin-phonon effects. Couple optical phonon modes are observed in the LT (23 K) antiferromagnetic phase and are strongly suppressed at room-temperature. Based on the DFT calculations,  optical phonons have been identified to arise from bond-bending modes of the Cr-S-Cr network, establishing the lattice origins of the observed excitations. In contrast to their absence at RT, the appearance of sharp, coherent optical phonon excitations at 23 K highlights a pronounced temperature dependence in electron–phonon coupling. This behavior suggests that at low temperatures, these phonons become coherent enough to couple effectively to the Cr 3$d$ electronic states, acting as local vibrational modes that modulate ligand fields. Their\sout{ disappearance} intensity decrease at higher temperatures likely results from significant anharmonic broadening or symmetry suppression due to thermal lattice fluctuations, rendering them \emph{dark} to RIXS despite increased thermal occupation. This transition indicates a structural or spin-driven distortion that enhances the sensitivity of the chromium electronic environment to specific lattice displacements, stabilized by magnetoelastic or symmetry-lowering effects unique to the magnetically ordered state of CrSBr. By resolving these coupled excitations, RIXS establishes itself as a sensitive tool for investigating layered magnetic semiconductors and the complex degrees of freedom within 2D vdW materials.

While the observed suppression of the low-energy RIXS feature between the magnetically ordered low-temperature phase and the paramagnetic room-temperature phase is consistent with the proposed spin-phonon coupling framework, the present measurements are limited to two temperatures. Consequently, the current data do not directly establish how the RIXS spectral weight evolves through the magnetic transition temperature. Nevertheless, the combined experimental observation, phonon calculations, and spin-renormalized electron-phonon RIXS analysis collectively identify spin-phonon coupling as the most consistent explanation for the observed temperature-dependent suppression of the low-energy spectral feature. Future temperature-dependent RIXS measurements across the magnetic phase transition would provide a more stringent test of this interpretation.

The current findings open a new window for understanding phonon contributions to magnetic anisotropy and provide a pathway toward controlling coupled excitations for future optoelectronic and spintronic applications. Furthermore, this research points to RIXS as a vital method for exploring electron–phonon interactions that may directly influence the excitonic and transport properties of next-generation materials. These findings provide a pathway towards understanding and controlling coupled excitations in vdW magnets for future optoelectronic and spintronic applications~\cite{yang2025multi}.

\vspace{0.1cm}
{\flushleft \textbf{Methods}}

{\flushleft \textbf{Sample Preparation.}} Large single crystals of CrSBr was synthesized by using the procedure explained by Scheie $et$ $al$ \cite{scheie2022spin}. Chromium (0.174 g, 3.35 mmol), sulfur (0.196 g,6.11 mmol), and CrBr$_3$ (0.803 g, 2.75 mmol) were loaded into a 12.7 mm O.D., 10.5 mm I.D. fused silica tube. The tube was evacuated to a pressure of $\sim$30 mtorr and flame sealed to a length of 20 cm. The tube was placed into a computer-controlled, two-zone, tube furnace. The source side was heated to 850 $^\circ$C in 24 h, allowed to soak for 24 h, heated to 950 $^\circ$C in 12 h, allowed to soak for 48 h, and then cooled to ambient temperature in 6 h. The sink side was heated to 950 $^\circ$C in 24 h, allowed to soak for 24 h, heated to 850 $^\circ$C in 12 h, allowed to soak for 48 h, and then cooled to ambient temperature in 6 h. The crystals were cleaned by soaking in a 1 mg mL$^{-1}$ of CrCl$_2$ aqueous solution for 1 h at ambient temperature. After soaking, the solution was decanted and the crystals were thoroughly rinsed with DI water and acetone. Residual sulfur residue was removed by washing with warm toluene.

{\flushleft \textbf{Experimental set-up.}} Cr $L_{2,3}$ XAS and RIXS experiments were performed at the I21 RIXS beamline of the Diamond Light Source (DLS) in the UK \cite{zhou2022i21}. The beamline provides a highly monochromatized, focused, tunable x-ray beam onto materials. The detection and energy-analysis of scattered x-rays is achieved using a spatially-resolved two-dimensional detector. The beam size at the sample is {$\sim$30 x 2} $\mu$m${^2}$ (Horizontal X Vertical) \cite{zhou2022i21}. The energy resolution of the beamline is controlled by using divergent variable line spacing gratings which can reach up to a maximum resolution of {$\sim$24} meV. The inset of Figure~\ref{XAS} shows the experimental geometry we used for the data acquisition. In $\pi$ ($\sigma$)-polarization measurements, the incident x-ray beam is polarized in the horizontal (vertical) scattering plane. We collected both XAS and RIXS data at low temperature (LT = 23 K) and at room temperature (RT = 300 K) for both $\pi$ and $\sigma$ polarizations. The spectrometer was operated with an experimental energy resolution of 24.5 meV at the Cr $L$-edge, unless otherwise specified. The general experimental procedure can be categorized into two main steps. First, XAS data were collected at the Cr $L$-edge, and then the incident x-ray beam energy was tuned into the XAS peak energies to collect the RIXS data. 

{\flushleft \textbf{Density Functional Theory (DFT) and phonon calculation.}} We perform first principles density functional theory calculations using Quantum Espresso (QE) \datta{\cite{giannozzi2009quantum,giannozzi2017advanced}}. These calculations utilized the Perdew-Burke-Ernzerhof (PBE) generalized gradient approximation (GGA)~\datta{\cite{perdew1996generalized}}. Optimized Norm-Conserving Vanderbilt (ONCV) pseudopotentials \cite{Hamann2013} (from PseudoDojo \cite{vanSetten2018}) were used. Calculations were performed both for a monolayer and bulk \csbr. In the case of the monolayer, a vacuum of 15 \(\text{\AA}\) was employed to mimic a monolayer setup, which was found to be sufficient to eliminate interactions between adjacent layers due to periodic boundary conditions. In the bulk configuration, the semiempirical Grimme's DFT-D2 van der Waals (vdW) correction was applied \cite{grimme2010consistent}. For both configurations, the lattice constants and atomic positions were relaxed such that the energy difference was less than \(7\times 10^{-8}\) Ry and the residual forces on the atoms were smaller than \(4\times 10^{-4}\) Ry/au . The kinetic energy cutoff for plane waves was set to 85 Ry. Note, DFT+U studies demonstrate that adding an effective Hubbard $U$ leads to an incorrect magnetic ground state for bulk \csbr~\cite{liu2024intralayer, li2025stacking,wang2022magnetic,shi2025controllable,henriquez2025strain}. Thus, we did not implement any Hubbard $U$ corrections. We constructed the bulk unit cell of the antiferromagnetic (AFM) configuration with two ferromagnetic layers, which are oppositely spin-polarized~\cite{lee2021magnetic}. We performed the calculations on a monolayer $3\times 3 \times 1$ supercell with a $k$-meshgrid of $6\times5\times1$ and a bulk $2\times 2 \times 2$ supercell with a $k$-meshgrid of $7\times6\times3$ in QE. To compute the phonon dispersion and the atom-projected phonon density of states (PDOS), we used the PHONOPY software package~\cite{phonopy-phono3py-JPCM,phonopy-phono3py-JPSJ}. Utilizing the same supercells as before, a mesh-grid sampling of $24\times24\times1$ (monolayer) and $28\times28\times12$ (bulk) was used to compute the PDOS in PHONOPY. Subsequently, the SMODES software suite \cite{stokes2007isotropy} in conjunction with the Phonon website (\url{https://tinyurl.com/phononwebsite}) was utilized to analyze the irreps and visualize the phonon vibrations. The data for the irreps are tabulated in Supplementary Material Table S1. 

{\flushleft \textbf{Acknowledgments}}

{\flushleft The} authors would like to acknowledge Diamond Light Source for providing beamtime under the proposal number MM33041, and the staff of I21 RIXS beamline for assistance with data acquisition. The Welch Foundation (grant number: E-0001) and the Texas Center for Superconductivity (TcSUH) supported work at the University of Houston. Part of this work was supported by the U.S. DOE, BES, under Award No. DE-SC0024332. The authors acknowledge support from the U.S. Air Force Office of Scientific Research and Clarkson Aerospace Corp. under Award FA9550-21-1-0460. Special thanks to the eXn group members at the University of Houston. B.B. and T.D. acknowledge funding support from the Augusta University Provost's office, the Student Research Program (SRP) of the Department of Medicine, Medical College of Georgia at Augusta University, and from the Center for Undergraduate Research and Scholarship (CURS) at Augusta University. B.B. and T.D. acknowledges Augusta University High Performance Computing Services (AUHPCS) for providing computational resources contributing to the results presented in this publication. B.B. and T.D. acknowledge that this work used Anvil CPU at Purdue University through allocation PHY240303 from the Advanced Cyberinfrastructure Coordination Ecosystem: Services \& Support (ACCESS) program, which is supported by U.S. National Science Foundation grants \#2138259, \#2138286, \#2138307, \#2137603, and \#2138296. B.B. and T.D. acknowledge helpful discussions on DFT with Turan Birol. The work at The University of Washington is supported by AFOSR FA9550-24-1-0004.

{\flushleft \textbf{Author Contributions}}

{\flushleft {J.P., C.P. and S.A. conducted the XAS/RIXS measurements. J.C., X. R. and X.X. synthesized and characterized the samples. The x-ray results were initially processed and analyzed by J.P. and C.P. J.P. and C.P. performed data simulations. B. B. and T.D. conducted DFT, phonon, and SMODES calculations. T.D. formulated the spin-phonon RIXS theory analysis. The overall results were collectively analyzed and discussed by all authors. J. P., T. D. and B.F. wrote the paper, incorporating input from all co-authors. Y.C. provided analysis advice and help with experiment planning. B.F. conceived and planned experiments, supervised the project and acquired funding to support the experimental work.}}

{\flushleft \textbf{Data Availability}}
The data that support the findings of this study are available from the corresponding authors upon reasonable request.

\bibliographystyle{naturemag}
\bibliography{References}

{\flushleft \textbf{Competing interests}}

{\flushleft All} authors declare no financial or non-financial competing interests. 

{\flushleft \textbf{Additional information}} 

{\flushleft Additional} information that supports this work is available online in the Supplementary Information.

{\flushleft \textbf{Correspondence}} and requests for materials should be addressed to Byron Freelon (experiment) and Trinanjan Datta (theory and DFT).

\clearpage
\onecolumngrid
    
\begin{center}
{\Large\textbf{Supplementary Information: RIXS Identification of Optical Phonon-Spin Coupling Effects in CrSBr}}\\[1em]

Jayajeewana N.~Ranhili,$^{1}$ 
Chamini Pathiraja,$^{1}$ 
Brody Brogdon,$^{2}$ 
John Cenker,$^{3}$ 
Xiaodong Xu,$^{3}$ 
Daniel Chica,$^{4}$ 
Xavier Roy,$^{4}$\\
Stefano Agrestini,$^{5}$ 
Mirian Garcia-Fernandez,$^{5}$ 
Ke-Jin Zhou,$^{5}$ 
Yi-De Chuang,$^{6}$ 
Trinanjan Datta,$^{2,*}$ 
and Byron Freelon$^{1,\dagger}$

\vspace{0.6em}

\textit{
$^{1}$Physics Department and Texas Center for Superconductivity, University of Houston, Houston, TX 77204, USA\\
$^{2}$Department of Physics and Biophysics, Augusta University, Augusta, GA 30912, USA\\
$^{3}$Department of Physics, University of Washington, Seattle, WA 98195, USA\\
$^{4}$Department of Chemistry, Columbia University, New York, NY 10027, USA\\
$^{5}$Diamond Light Source, Harwell Campus, Didcot, OX11 0DE, UK\\
$^{6}$Lawrence Berkeley National Laboratory, Berkeley, CA 94720, USA
}

\vspace{0.6em}

\end{center}

\vspace{1em}

\section{$\mathrm{\textbf{CrSBr Crystal Structure}}$}

\begin{figure*}[hbtp]
   \centering
    \includegraphics[width=1\textwidth]{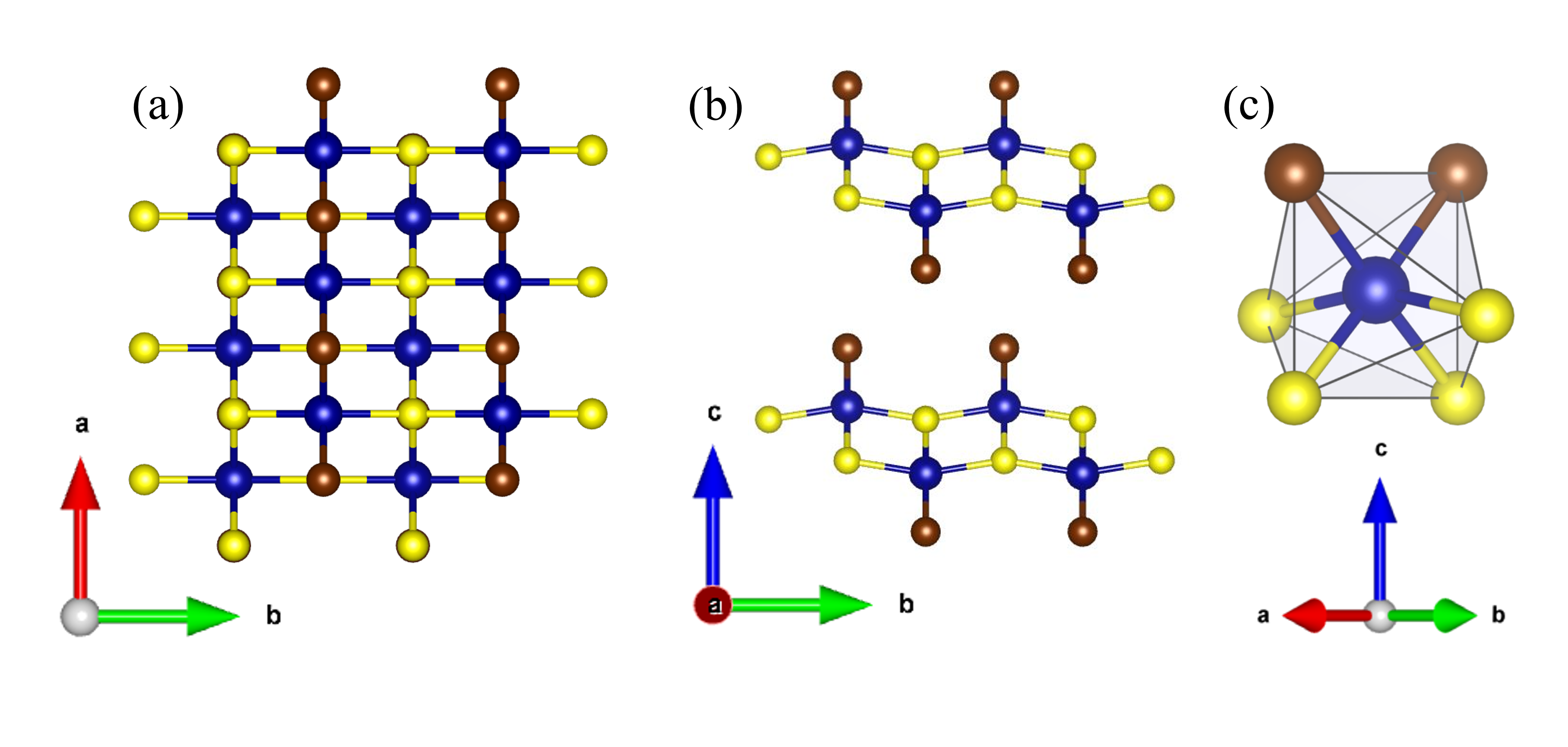}
    \caption{(a) The bond distribution of CrSBr along $a$ and $b$ directions. Blue, yellow and brown represents Cr, S and Br atoms respectively. (b) The stacking order of CrSBr layers along the $c$-axis. (c) The distorted octahedron. Crystal structure was plotted using VESTA. A given layer, which comprises of two chromium atoms (two blue spheres), have ferromagnetic intra-layer exchange. But, inter-layer coupling is (weak) A-type  antiferromagnetic below the N\'{e}el temperature.}
    \label{CrSBr_UnitCell}
\end{figure*}

Chromium sulfur bromide (CrSBr) is composed of rectangular monolayer lattices stacked along the crystallographic \textit{c} axis and held together by vdW interactions \cite{ziebel2024crsbr}. Each monolayer is composed of edge-sharing distorted octahedra ~\cite{boix2022probing}, in which Cr$^{3+}$ cations are surrounded by four S$^{2-}$ and two Br$^{-}$ anions \cite{wang2023magnetically, klein2022control}. The bond distribution along the $a$ and the $b$ axes, the stacking order along the $c$ axis, and the distorted octahedron are shown in Figure \ref{CrSBr_UnitCell}. The atomic bonding along the $b$-axis involves Cr and S atoms. Sulfur atoms bridge the Cr ions through the formation of Cr-S-Cr chains. These Cr-S-Cr bonds are crucial for mediating magnetic interactions \cite{long2023ferromagnetic}.

\section{$\mathrm{\textbf{XAS of $\sigma$-polarization}}$}
\begin{figure}[!ht]
     \centering
    \includegraphics[width=0.75\textwidth]{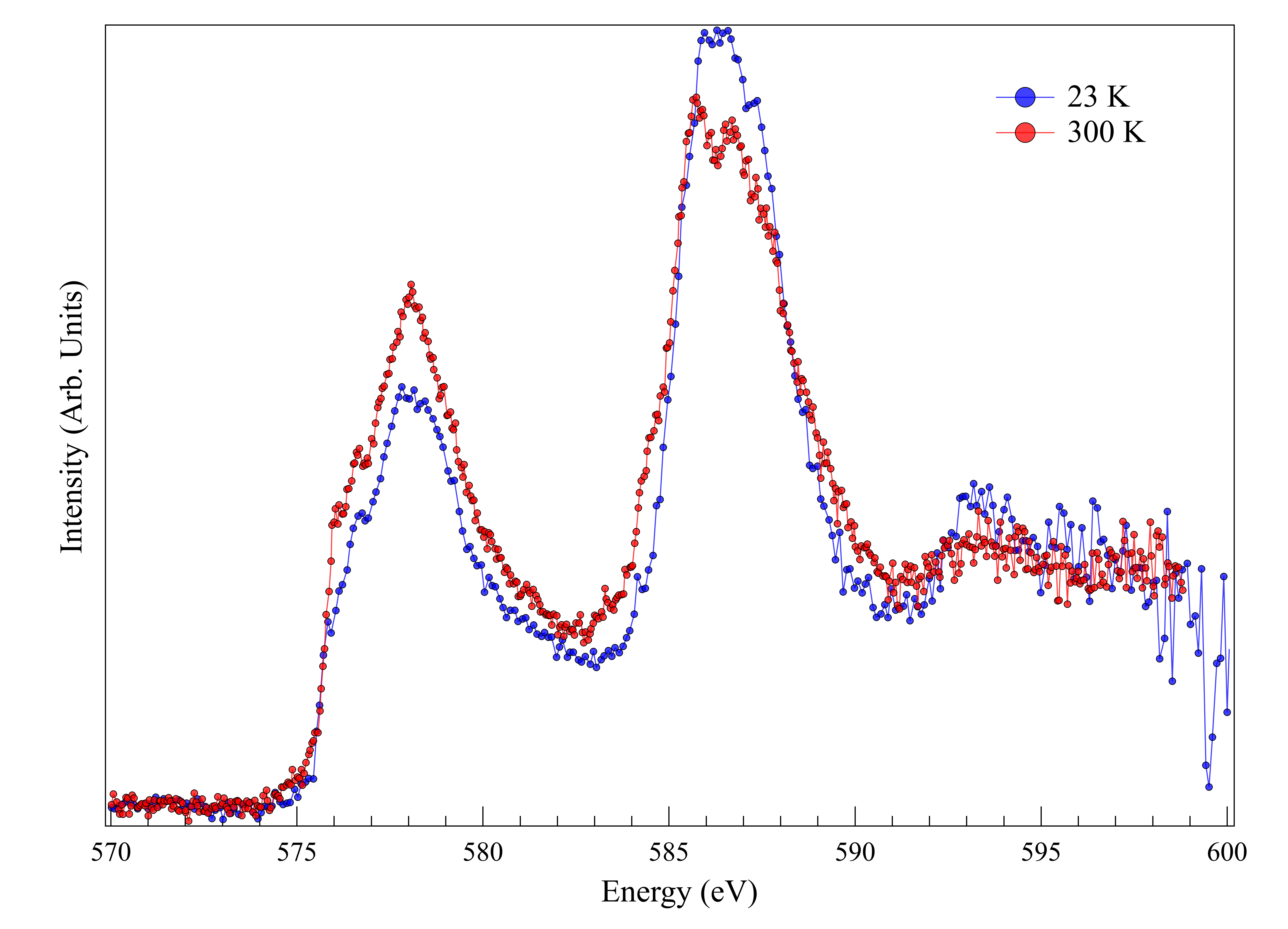}
    \caption{XAS spectra collected at 23 K (LT, blue) and 300 K (RT, red) under $\sigma$-polarization in total fluorescence yield (TFY) mode. The $L_3$ peak energy is near 577.8 eV, which is close to the $\pi$-polarization data.} 
    \label{XAS_sigma}
\end{figure}

Figure \ref{XAS_sigma} shows the XAS spectra at low-temperature (LT) and room temperature (RT) collected under $\sigma$-polarization in total fluorescence yield (TFY) mode. The plot has similar peak energies as in the $\pi$-polarization spectra.

\section{$\mathrm{\textbf{Quasi Elastic Peaks (QEPs) along the $a$ and $b$ axes of the Sample}}$}

\begin{figure}[hbtp]
     \centering
    \includegraphics[width=1\textwidth]{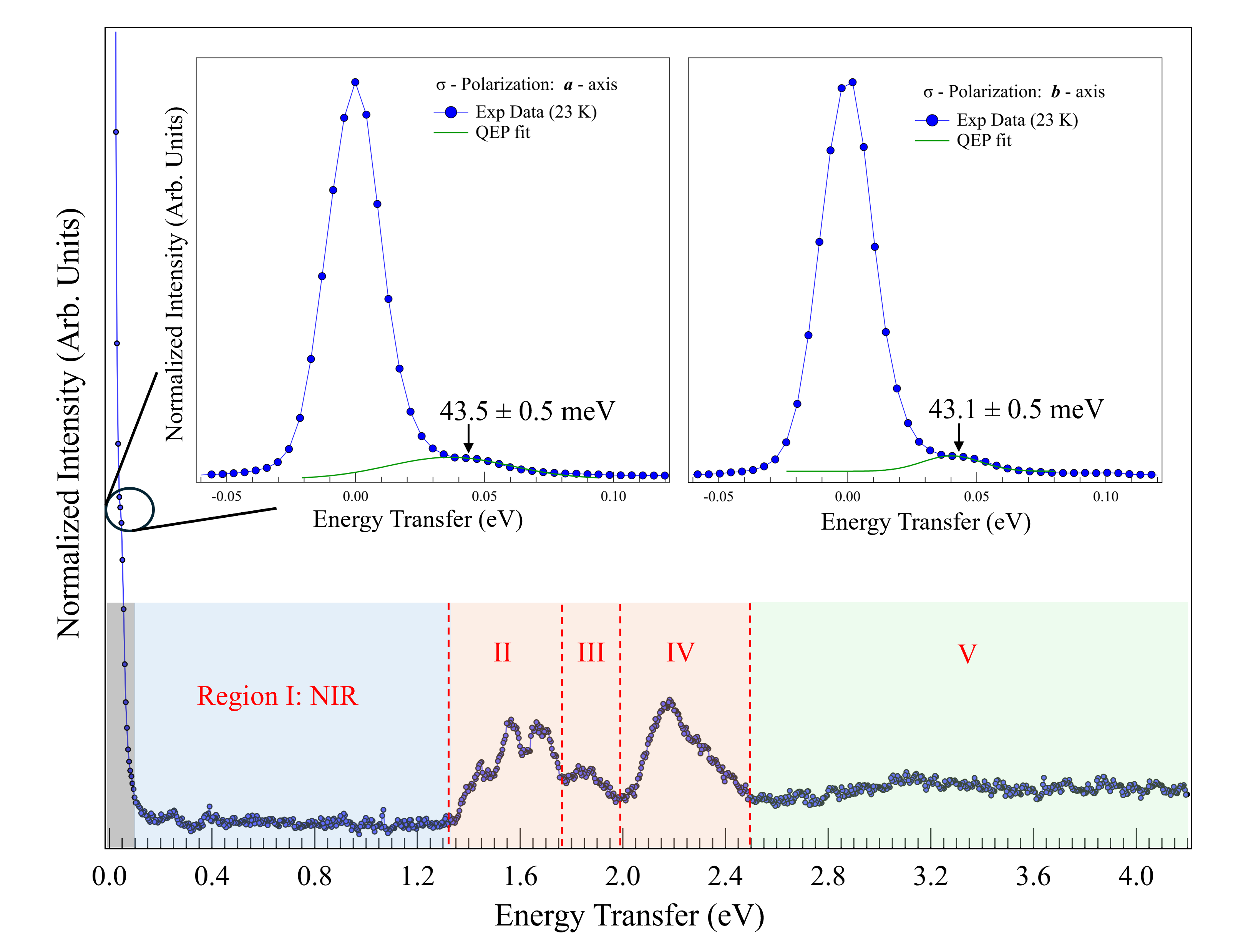}
    \caption{RIXS spectrum of bulk CrSBr collected under $\sigma$-polarization at the $\Gamma$ point at 23 K (LT). The spectrum is normalized at the elastic peak (EP) and divided into several regions to highlight the various observed excitations, which are explained in the main text. Two insets show the appearance of quasi-elastic peak (QEP) along the $a$ and $b$ axes respectively near to the EP in the low-temperature phase. QEP energy along each axis is 43.5 $\pm$ 0.5 meV and 43.1 $\pm$ 0.5 meV, respectively.}
    \label{QEP_Fig}
\end{figure}

The $\sigma$-polarization RIXS data were used to calculate the QEP energies along the $a$ and $b$ axes of the sample. Figure \ref{QEP_Fig} shows the elastic-peak-normalized RIXS spectrum collected under $\sigma$-polarization at the $\Gamma$ point at 23 K (LT). Quasi elastic peak energies along the $a$ and $b$ axes are 43.5 $\pm$ 0.5 meV and 43.1 $\pm$ 0.5 meV, respectively \cite{mudiyanselage2025determination}. 

\section{$\mathrm{\textbf{Phonon Mode Snapshots and Symmetry Analysis}}$}
\begin{figure*}[hbtp]
    \centering
    \includegraphics[width=0.87\linewidth]{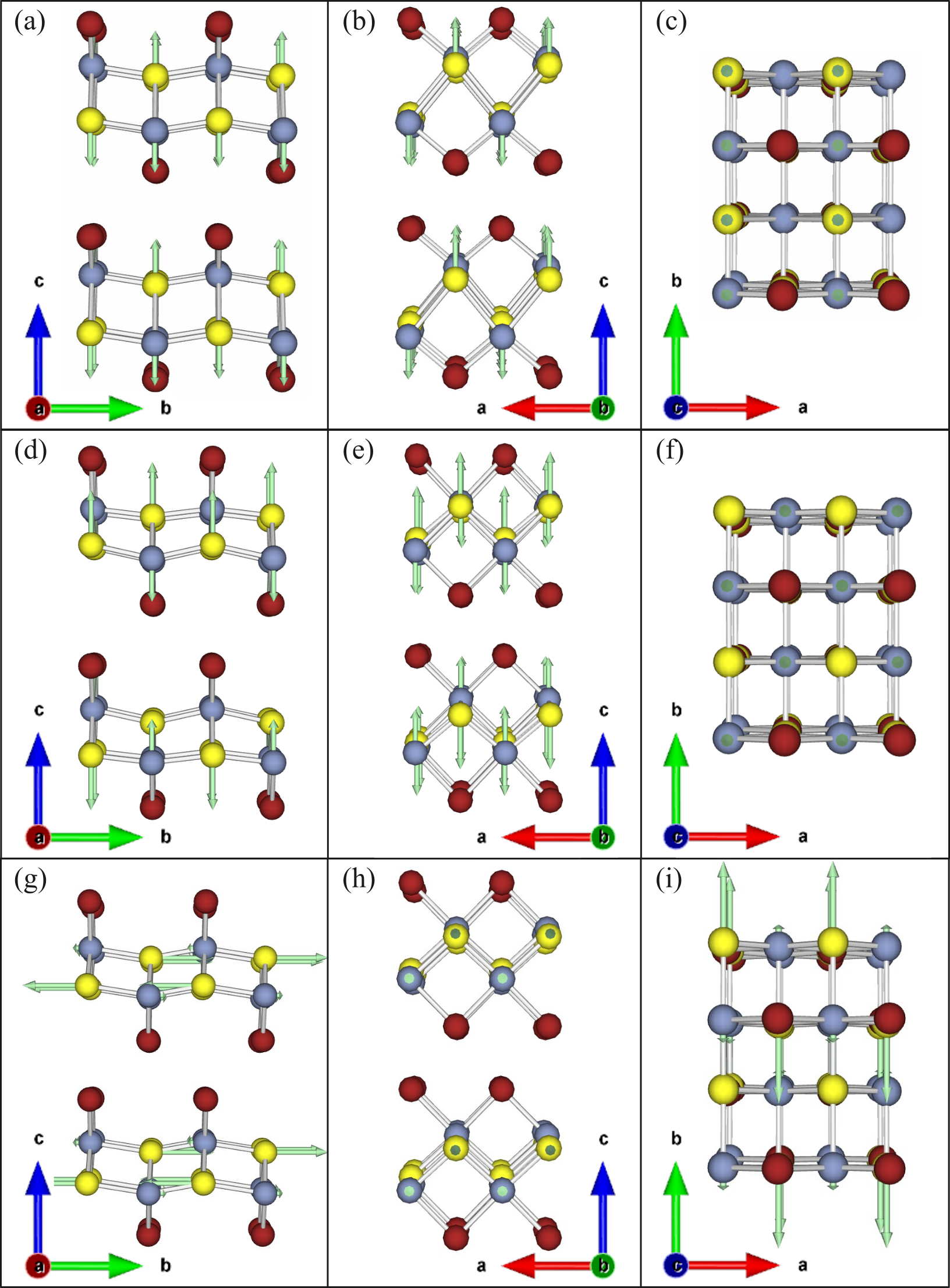}
    \caption{Lattice oscillation snapshots of the six highest optical phonon modes of bulk \csbr~created using the Phonon Website~\cite{PhononWebsite}. Blue, yellow and red spheres represent the Cr, the S and the Br atoms respectively. The green arrows (or dots) on the spheres indicate movements of the individual atoms at a specific time snapshot. The first, the second, and the third column represents oscillations in the $a$, $b$, and $c$ directions (see coordinate axes), respectively. The 31$^{\text{st}}$ and the 32$^{\text{nd}}$ phonon modes of $A_g$ symmetry are shown in panels (a)-(c). The 33$^{\text{rd}}$ and the 34$^{\text{th}}$ modes of $B_{1u}$ symmetry are shown in panels (d)-(e). The last row stands for the 35$^{\text{th}}$ and the 36$^{\text{th}}$ modes of $B_{3g}$ symmetry.}
    \label{fig:figS5}
\end{figure*}

The direction of the atomic movements where deduced using the Phonon website \cite{PhononWebsite}. The lattice displacement screenshots are shown in FIG.~\ref{fig:figS5}. We utilized SMODES \cite{stokes2007isotropy} to compute the irreducible representations and description of the atomic motion in the unit cell of \csbr. Table~\ref{tab:tab1} below summarizes the results for bulk \csbr. 

\begin{table*}[hbtp]
\centering
\begin{center}
\begin{tabular}{ |c|c|c|c| } 
\hline
~~Optic band label~~ & ~~Energy (meV,~THz)~~ &~~Symmetry~~ &~~Description of motion  \\
\hline
36 & 44.056, 10.653 & $B_{3g}$ & Movement along b, S-S \& Cr-Cr~out of phase \\ \hline
35 & 44.049, 10.651 & $B_{3g}$ & Movement along b, S-S \& Cr-Cr~out of phase \\ \hline

34 & 43.264, 10.461 & $B_{1u}$ & Movement along c, S-S \& Cr-Cr~ in phase  \\ \hline
33 & 42.416, 10.256 & $B_{1u}$ & Movement along c, S-S \& Cr-Cr~ in phase \\ \hline

32 & 40.903, ~9.891 & $A_g$  & Movement along c, S-S \& Cr-Cr~ out of phase\\ \hline
31 & 40.863, ~9.881 & $A_g$  & Movement along c, S-S \& Cr-Cr~ out of phase \\ \hline

\hline
\end{tabular}
\caption{The six top most optic bands, corresponding energies, irreducible representation, and description of the atomic displacement for bulk \csbr~at the $\Gamma$ point Note, movement along a specific cartesian direction implies that the atoms are displaced either in-phase or out-of-phase along these axes.}  
\label{tab:tab1}
\end{center}
\end{table*}

\section{$\mathrm{\textbf{Spin-Phonon Frequency Renormalization of CrSBr}}$}
We derive the spin-renormalized phonon frequency $\omega_\mathrm{r}$ and spring constant $\kappa_\mathrm{r}$ for a normal mode $Q$ from a microscopic model where the exchange coupling depends on $Q$ \cite{spphn1,spphn2,BirolPhysRevB.93.134425}. We derive an expression for the observable spin-renormalized phonon frequency $\omega_\mathrm{r}^2=\kappa_\mathrm{r}/M$, where $M$ is the mass of the atom. Consider a single phonon coordinate $Q$ (bare spring constant $\kappa_0=M\omega_0^2$, and $\omega_0$ is the bare resonant frequency) coupled to a spin system through the phonon coordinate dependent exchange coupling $J_{ij}(Q) > 0$, where $i$ and $j$ represent spin-sites. We write the spin-phonon Heisenberg Hamiltonian $H(Q)$ as \begin{equation}
\label{eq:hq}
H(Q) = \frac{P^2}{2M} + \frac{1}{2}\kappa_0 Q^2
\;-\;\sum_{\{ij\}} J_{ij}(Q)\, \mathbf S_i\!\cdot\!\mathbf S_j,
\end{equation}
where $P$ is the canonical momentum conjugate to $Q$. The third term represents the Heisenberg type spin-phonon coupling Hamiltonian. Here $\mathbf S_i$ is the spin operator at site $i$. The summation index pair $\{ij\}$, where $i$ and $j$ represent site indices, is over various nearest-neighbor $(nn)$ bonds. We consider the dominant intralayer exchange terms for \csbr~(up to the third neighbor). Further neighbor exchange or interlayer interactions are ignored since they are weaker in strength~\cite{Bo2023}. To consider the effects of lattice vibrations (phonons) on the exchange constant $J_{ij}(Q)$ we expand the magnetic interaction about $Q=0$ to have\begin{equation}
\label{eq:jq}
J_{ij}(Q) = J_{ij}^{(0)} + J'_{ij}\,Q + \frac{1}{2}J''_{ij}\,Q^2 + \mathcal{O}(Q^3),
\end{equation}
where the primes denote derivatives of $J(Q)$ with respect to $Q$. Next, we insert Eq.~\eqref{eq:jq} into \eqref{eq:hq}, group the $Q$ dependent terms, order-by-order, up to second order to rewrite the Hamiltonian as \begin{equation}H(Q) = \frac{P^2}{2M} + \frac{1}{2}\kappa_0 Q^2
- \sum_{\{ ij\}} J_{ij}^{(0)}\, \mathbf S_i\!\cdot\!\mathbf S_j 
- Q \sum_{\{ ij\}} J'_{ij}\, \mathbf S_i\!\cdot\!\mathbf S_j
- \frac{Q^2}{2}\sum_{\{ ij\}} J''_{ij}\, \mathbf S_i\!\cdot\!\mathbf S_j + \mathcal{O}(Q^3).
\end{equation}
To proceed further, we will assume that the time-scale over which the phonon dynamics evolves is much slower compared to that of the magnon dynamics. Because the spins fluctuate at a much faster rate than phonons, the phonon field cannot respond to instantaneous spin configurations; it only feels their time-averaged (thermal) effect. So, within this approximation, we can replace the Heisenberg spin interaction correlator with a mean field average, that is $C_{ij}(T) \equiv\langle\mathbf{S_i}\cdot\mathbf{S_j}\rangle$, where $T$ is the temperature. Next, we define a potential energy function $V(Q)=\tfrac12 \kappa_{\mathrm{r}} Q^2$, where the renormalized spring constant $\kappa_{\mathrm{r}} = \kappa_0 - \sum\limits_{\{ ij\}} J''_{ij} \, C_{ij}(T)$. Thus, the renormalized phonon frequency $\omega_{\mathrm{r}}$ is given by \begin{equation}\label{eq:omr} \omega_{\mathrm{r}}^2 = \frac{\kappa_{\mathrm{r}}}{M}
= \omega_0^2 - \frac{1}{M}\sum_{\langle ij\rangle} J''_{ij}\, C_{ij}(T).
\end{equation} The expansion process generates a linear contribution $-FQ$, where $F=\sum\limits_{\{ ij\}} J'_{ij} C_{ij}(T)$. But, this simply shifts the equilibrium position to $Q^{*}=F/\kappa_{\mathrm{r}}$. Note, this term does not renormalize the phonon energy modes and is not important for our consideration. 

Lattice vibration snap shots from phonon density functional theory (DFT) calculations (see Supplementary Figure~S4) indicate that bond-bending modes (bands 31 - 34 where both the Cr and S move, Br has negligible motion) or stretching modes (35 and 36 where S displaces primarily, Br has negligible motion) can give rise to phonon vibrations. The $\sim$43 meV RIXS peak suggests that we can focus on the bond-bending mode of the Cr--S--Cr superexchange pathway in \csbr~. Let us consider this angle to be $\theta$. Thus, the exchange constant $J_{s} = J_{s}(\theta(Q))$, where $s = \{ij\}_n$ represents the intralayer $(nn)$: $n=1$ (first), $n=2$ (second), and $n=3$ (third). For small displacements and assuming $\theta(Q) \approx \theta_0 + \lambda_s Q$ (linear variation from the equilibrium bond angle $\theta_0$), we have  (applying the chain rule) $J_{s}^{\prime\prime} = K_s\lambda_s^2$ for each $nn$ coordination where $K_s=\partial^2J_s/\partial^2\theta$ and $\lambda_s = \partial \theta/\partial Q$. The first factor $K_s$ represents the magnetic stiffness in response to bond-angle distortions. The second factor $\lambda_s$ is a geometric effect of bond angle variation on the phonon normal coordinate $Q$. 

To proceed further, we need to conceptually understand how $K_s$ and $\lambda_s$ will affect the phonon frequency renormalization process. To explore this variation we consider the magnetic phase transitions that the \csbr~compound undergoes as the experimental data is probed from room temperature (RT) to 23 K. As outlined in the main text, the magnetic ordering of \csbr~can be thermally driven. Near 146 K the material develops an in-plane ferromagnetic order~\cite{rudenko2023dielectric,ziebel2024crsbr}. However, below approximately 132 K, weak antiferromagnetic coupling between the layers develops concomitantly with strong intra-layer ferromagnetic interaction to stabilize a long-range $A$-type antiferromagnetic order ~\cite{goser1990magnetic,rudenko2023dielectric,ziebel2024crsbr}. From a minimal analysis viewpoint, we can apply the Goodenough-Kanamori-Anderson rules~\cite{GoodenoughPhysRev.100.564,Kanamori195987,AndersonPhysRev.115.2} modeling of the exchange constant on the Cr--S--Br bond angles to write $J_s(\theta) = J^{(s)}_{AFM}\cos^2(\theta) - J^{(s)}_{FM}\sin^2(\theta)$ where the subscripts on the exchange constants have their usual meaning. This implies $ K_s = -2\,(J_{\mathrm{AF}}^{(s)}+J_{\mathrm{F}}^{(s)})\,\cos\big(2\theta_{s,0}\big)$, where $\theta_{s,0}$ is the equilibrium angle for the given shell $s$. The phonon eigenvector sets the $\lambda_s$ parameter based on the nature of the atomic motion. The sign of $K_s$ is established by the equilibrium angle $\theta_{s,0}$. Close to $180^\circ$, $K_s<0$ (antiferromagnetic), but near $90^\circ$ $K_s>0$ (ferromagnetic). Inserting this into Equation~\eqref{eq:omr} yields the renormalized frequency in terms of these geometric and magnetic quantities relevant for a particular normal mode. Note, the sign for $\lambda_s$ is irrelevant since this quantity will enter as a square value in the final expression for frequency renormalization.

Next, we utilize the above expressions to obtain the renormalized phonon frequency $\omega_r$ for \csbr~. Energy values in Table S1 indicate that the 34$^{\text{th}}$ band is at 43.264 meV. This value is close to the $a$ axis RIXS peak energy (compare with 43.5 meV). Hence, as a concrete example, and to demonstrate our spin-phonon coupled RIXS hardening effect, we choose the bond bending mode as mentioned earlier in the main text and the supplementary. In our example, the Cr--Cr and the S--S sublattices move out of phase. Therefore, only the Cr--S--Cr bond-bending interactions contribute to the spin--phonon coupling for these modes. Thus, the renormalized phonon frequency $\omega_{\mathrm{r}}$ is given by 

\vspace{-0.25cm}
\label{eq:omcsb}
\begin{equation}
\omega_{\mathrm{r}}^2=\omega_0^2+\frac{\Delta\kappa}{M},
\qquad
\Delta\kappa \equiv \kappa_{\mathrm{r}}-\kappa_0
= -2\Big[
(\,J_{1a}''\,C_{1a}
+\,J_{1b}''\,C_{1b}
+2\,J_{2}''\,C_{2}
+\,J_{3a}''\,C_{3a}
+\,J_{3b}''\,C_{3b})
\Big] ,
\end{equation}

where we suppressed the temperature argument in $C_{\{ij\}}(T)$ for convenience. Depending on the $\{i,j\}$ $nn$ we perform the $nn$ summation for the \csbr~compound to identify $J_s''$. Utilizing Eq.~\eqref{eq:omr}, we can expand the summation over the $nn$ bonds as: $1a$: first neighbors along $\pm a$ (two bonds), exchange $J_{1a}$, correlator $C_{1a}$; $1b$: first neighbors along $\pm b$ (two bonds), exchange $J_{1b}$, correlator $C_{1b}$; $2$: diagonal second neighbors $\,\pm a \pm b$ (four bonds), exchange $J_2$, correlator $C_2$; $3a$: third neighbors along $\,\pm 2a$ (two bonds), exchange $J_{3a}$, correlator $C_{3a}$; and $3b$: third neighbors along $\,\pm 2b$ (two bonds), exchange $J_{3b}$, correlator $C_{3b}$. The $nn$ coordination numbers $(2,2,4,2,2)$ appear as coefficients in the $\Delta\kappa$ expression, above. Next, we substitute $J_s''$ with $K_s\lambda_s^2$ to obtain the renormalized phonon frequency $\omega_r$ as
\vspace{-0.25cm}
\begin{equation}
\label{eq:wrfull}
\omega_{\mathrm{r}}^2
=\omega_0^2
-\frac{2}{M}\Big[
(K_{1a}\lambda_{1a}^2 C_{1a}
+K_{1b}\lambda_{1b}^2 C_{1b}
+2K_{2}\lambda_{2}^2 C_{2}
+K_{3a}\lambda_{3a}^2 C_{3a}
+K_{3b}\lambda_{3b}^2 C_{3b})
\Big].
\end{equation}
The planar interactions are all ferromagnetic and dominate the physics.  These correlators are all positive. While it maybe possible to evaluate term-by-term the exact values of the magnetic stiffness for each of bond contributions in Eq.~\eqref{eq:omgC}, we will proceed with a more simplistic model. To keep the discussion conceptual and the analysis tractable, we consider the most dominant FM interaction. Thus, for discussion purposes Eq.~\eqref{eq:wrfull} can be rewritten as 
\begin{equation}
\label{eq:omgC}
\omega_{\mathrm{r}}^2
=\omega_0^2 - \sum\limits_s \mathrm{p}_s C_s \approx \omega_0^2 - \mathrm{p}_lC_{l},
\end{equation}
where $\mathrm{p}_l$ and $C_l$ represents the leading order term in the renormalized frequency expression (the expression quoted in the main text). Note, $\mathrm{p}_l = K_l\lambda^2_l/M $ is the weight for the most dominant bond contribution (in our case from the $K_2$ term based on neutron scattering fit)~\cite{Bo2023}. For an intralayer ferromagnetic channel $K_l > 0$. Furthermore, $C_l(T)>0$ for planar spin-spin correlation. This equation suggests that at low temperatures, the phonon frequency is softened (since $C_l =1$). However, with increasing temperature as the ferromagnetic correlations weaken, $\omega_{\mathrm{r}} \rightarrow \omega_0$ thereby causing the phonon frequency to harden. This hardening has an impact on the electron-phonon coupled RIXS spectrum. Based on electron-phonon coupled RIXS theory~\cite{ament2011determining}, we can infer that dimensionless electron-phonon coupling $g$ is inversely proportional to the square of the phonon frequency (that is $g\propto 1/\omega^2_r$). Thus, as the frequency hardens, the coupling $g$ weakens. Computing the RIXS spectrum for various coupling values, see Figure~4 (in the main text), we notice that the RIXS spectrum is suppressed with smaller values of $g$. Thus, hardening of phonon frequency with increasing temperature explains why the RIXS peaks get suppressed at room temperature.

\end{document}